\documentclass[preprint2,10.5pt]{aastex}
\begin{document}

\title{\large{\rm{ALESSI 95 AND THE SHORT PERIOD CEPHEID SU CASSIOPEIAE}}}
\author{D.~G. Turner$^{1,9}$, D.~J. Majaess$^{1,2,9}$, D.~J. Lane$^{1,2}$, D.~D. Balam$^3$, W.~P. Gieren$^4$ \\ J. Storm$^5$, D.~W. Forbes$^6$, R.~J. Havlen$^{7,10}$, B. Alessi$^8$}
\affil{$^1$Saint Mary's University, Halifax, NS, Canada}
\affil{$^2$The Abbey Ridge Observatory, Stillwater Lake, NS, Canada}
\affil{$^3$Dominion Astrophysical Observatory, Victoria, BC, Canada}
\affil{$^4$Universidad de Concepci\'{o}n, Concepci\'{o}n, Chile}
\affil{$^5$Leibniz-Institut f\"{u}r Astrophysik Potsdam (AIP), Potsdam, Germany}
\affil{$^6$Sir Wilfred Grenfell College, Memorial University, Corner Brook, Newfoundland, Canada}
\affil{$^7$307 Big Horn Ridge, NE, Albuquerque, NM, 87122, U.S.A.}
\affil{$^8$Universidade de S\~{a}o Paulo, S\~{a}o Paulo, Brazil}
\affil{$^9$Guest Investigator, Dominion Astrophysical Observatory}
\affil{$^{10}$Visiting Astronomer, Kitt Peak National Observatory}
\email{turner@ap.smu.ca}

\begin{abstract}
The parameters for the newly-discovered open cluster Alessi 95 are established on the basis of available photometric and spectroscopic data, in conjunction with new observations. Colour excesses for spectroscopically-observed B and A-type stars near SU Cas follow a reddening relation described by $E_{U-B}/E_{B-V}=0.83+0.02E_{B-V}$, implying a value of $R=A_V/E_{B-V}\simeq2.8$ for the associated dust. Alessi 95 has a mean reddening of $E_{B-V}{\rm (B0)}=0.35\pm0.02$ s.e., an intrinsic distance modulus of $V_0-M_V=8.16\pm0.04$ s.e. ($\pm0.21$ s.d.), $d=429\pm8$ pc, and an estimated age of $10^{8.2}$ yr from ZAMS fitting of available {\it UBV}, CCD {\it BV}, {\it NOMAD}, and 2MASS {\it JHK}$_s$ observations of cluster stars. SU Cas is a likely cluster member, with an inferred space reddening of $E_{B-V}=0.33\pm0.02$ and a luminosity of $\langle M_V \rangle = -3.15 \pm0.07$ s.e., consistent with overtone pulsation ($P_{\rm FM}=2^{\rm d}.75$), as also implied by the Cepheid's light curve parameters, rate of period increase, and {\it Hipparcos} parallaxes for cluster stars. There is excellent agreement of the distance estimates for SU Cas inferred from cluster ZAMS fitting, its pulsation parallax derived from the infrared surface brightness technique, and {\it Hipparcos} parallaxes, which all agree to within a few percent.
\end{abstract}

\keywords{Galaxy: open clusters and associations: individual: Alessi 95---stars: variables: Cepheids---stars: individual: SU Cas.}

\section{{\rm \footnotesize INTRODUCTION}}

When \citet{vd66} compiled a list of relatively bright stars visible in the National Geographic-Palomar Observatory Sky Survey (POSS) that are associated with reflection nebulosity, he noted that the distance and luminosity for one such star, the $1^{\rm d}.949$ Cepheid SU Cas, might be estimated using spectroscopic and photometric observations of the nearby B-type stars HD~17138 and HD~17443, which appear to illuminate a portion of the same dust complex. His discovery was very important, given the complete lack in existing surveys of Galactic clusters lying within several degrees of the Cepheid that might serve as distance indicators. Curiously, the field of SU Cas was not surveyed in van den Bergh's earlier search of the POSS for previously-undetected star clusters \citep{vd57}.

\begin{figure*}
\begin{center}
\includegraphics[width=10.2cm]{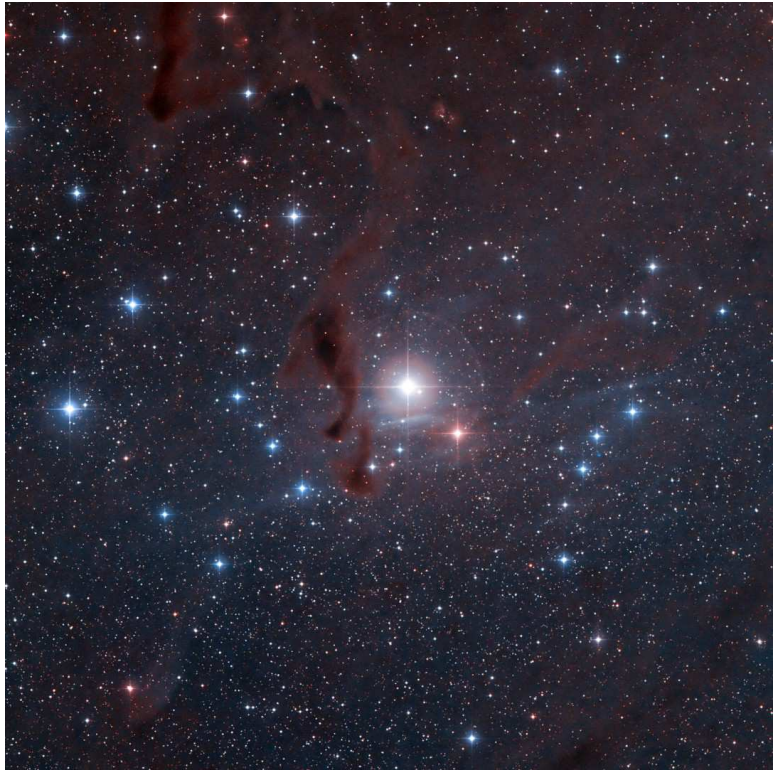}
\end{center}
\caption{\small{A composite colour $60^{\prime}\times60^{\prime}$ image of Alessi 95 centred on 2000.0 co-ordinates 02:52:15.1, +68:53:19, compiled by Noel Carboni from Palomar Observatory Sky Survey-2 blue, red, and near infrared images. SU Cas is the bright star near the centre of the image; the red star southwest of it is the K0 III star TE 10.}}
\label{fig1}
\end{figure*}

The sparse group of B stars was subsequently designated as Cas R2, and used by \citet{rv70} with other R-associations to map local spiral structure in the Galaxy. An initial estimate for the distance to the dust complex by \citet{ra68} yielded a value of $d=310 \pm30$ pc from spectroscopic distance moduli for the three B-type stars HD~17327, HD~17443, and HD~17706, and the bright M giant HD~23475. A later study by \citet{sc78} using Str\"{o}mgren and H$\beta$ photometry implied that HD~17327 and HD~17443 were closer than HD~17706, at a mean distance of $\sim 275$ pc, versus $\sim 440$ pc for the latter. Schmidt's result found support in an independent earlier survey by \citet{hv71} of B-type stars lying within $\sim 6^{\circ}$ of SU Cas. Only a few were found to lie at distances comparable to that expected for SU Cas, provided it was assumed to be either a fundamental mode or overtone pulsator. A more detailed study of Cas R2 stars by \citet{te84} included an additional star, HD~16893, associated with reflection nebulosity and an A-type companion to HD~17327, yielding a small group of five stars, with five additional candidates, lying at an estimated distance of $261 \pm 21$ pc. It was noted that background B-type stars were plentiful in the field, with most, including HD~17706, congregating at distances of $401 \pm 38$ pc.

The inferred luminosity for SU Cas requires an estimate of its reddening, but the immediate field of the Cepheid was surveyed by \citet[][see also \citet{te84}]{tu84}, and the field reddening appeared to be fairly well established, although with some dependence on localized dust obscuration. Several of the surveyed companions to SU Cas were found to have intrinsic distance moduli $V_0$--$M_V$ near 8.0, but there were no suspicions at the time that the Cepheid might be embedded in an anonymous open cluster. The situation changed dramatically a few years ago when Bruno Alessi discovered a sparse cluster of stars surrounding SU Cas and centred at J2000.0 co-ordinates 02:52:15.1, +68:53:19 \citep{al06}, as shown in Fig.~\ref{fig1}. The cluster, designated Alessi 95, has nuclear and coronal radii of $24^{\prime}$ and $50^{\prime}$, respectively, and was discovered from an analysis of on-line databases \citep*[see][]{al03,kr06}. The colour image (Fig.~\ref{fig1}) emphasises the large number of blue stars (B and A-type) dominating the central regions of the cluster, and also provides a clear picture of the relationship between SU Cas and its reflection nebula with the surrounding dust.

This study was initiated in order to provide additional information about Alessi 95 --- its reddening, distance, and age --- based upon available observational material, and to discuss what it reveals about SU Cas. For example, the revised {\it Hipparcos} parallax of $2.53 \pm0.32$ mas for SU Cas \citep{vl07} implies a distance of $395 \pm50$ pc to the Cepheid, consistent with intrinsic distance moduli near 8.0 found for several of its optical companions. It may therefore be possible to establish the distance to SU Cas via trigonometric, open cluster, and pulsation parallaxes, which would solidly anchor the short-period end of the Galactic calibration of the Cepheid period-luminosity relation and strengthen the relationship established by \citet{tu10} from open clusters and associations.
 
 \section{{\rm \footnotesize OBSERVATIONAL DATA AND ANALYSIS}}

An informative set of observations for bright B-type stars in the field of SU Cas was obtained four decades ago by \citet{hv71}, who made photoelectric {\it UBV} measures for 31 of 33 stars lying within $\sim 6^{\circ}$ of the Cepheid on two nights in September 1968 with the 0.9m telescope at Stewart Observatory. Spectroscopic observations for 25 of the stars (up to 3 spectrograms) were obtained at a dispersion of 63 \AA$\!$ mm$^{-1}$ with the No.~1, 0.9m telescope at Kitt Peak National Observatory (KPNO) during the summer of 1968. The spectra were measured for radial velocity and classified by \citet{hv71} using the facilities of KPNO, and also remeasured for radial velocity using the PDS microdensitometer at the University of Toronto, and reclassified by the lead author, for use in the study by \citet{te85}. Part of the latter study included Str\"{o}mgren and H$\beta$ photometry for the stars obtained by Forbes using the automatic photometer on the 0.76m telescope of the Behlen Observatory at the University of Nebraska, the same facility used in the study by \citet{sc78}.

\begin{table} \scriptsize 
\caption[]{Photometric and spectroscopic data for stars in the region of SU Cas.}
\label{tab1}
\begin{center}
\begin{tabular}{@{\extracolsep{-1.5mm}}lccclcc}
\hline
Star &{\it V} &{\it B--V} &{\it U--B} &Sp.T. &$E_{B-V}$ &$E_{U-B}$ \\
\hline
HD~11529 &4.98 &--0.09 &--0.42 &B7~IV &0.03 &0.05 \\
HD~11744 &7.82 &+0.38 &--0.31 &B3~III: &0.58 &0.47 \\
HD~12301 &5.61 &+0.38 &--0.28 &B7~Ib &0.44 &0.30 \\
HD~12509 &7.09 &+0.34 &--0.53 &B1~II &0.58 &0.46 \\
HD~12567 &8.32 &+0.39 &--0.54 &B0.5~III &0.67 &0.50 \\
HD~12882 &7.58 &+0.37 &--0.50 &B6~Iae &0.45 &$\cdots$ \\
HD~13590 &8.01 &+0.38 &--0.37 &B2~IIIe &0.62 &0.53 \\
HD~13630 &8.80 &+0.36 &+0.06 &B8~V &0.45 &0.36 \\
HD~14010 &7.14 &+0.60 &--0.10 &B8~Iab &0.63 &0.43 \\
HD~14863 &7.76 &+0.06 &--0.40 &B5~V &0.22 &0.18 \\
HD~14980 &9.10 &+0.42 &+0.02 &A0 III: &0.42 &$\cdots$ \\
HD~15472 &7.88 &+0.06 &--0.61 &B3~Ve &0.26 &$\cdots$ \\
HD~15727 &8.25 &+0.50 &--0.20 &B3~III:nn &0.70 &0.58 \\
HD~16036 &8.19 &+0.44 &+0.27 &B9~Vn &0.50 &0.45 \\
HD~16393 &7.59 &+0.04 &--0.31 &B7~Vnn &0.16 &0.11 \\
HD~16440 &7.89 &+0.75 &+0.10: &B3~Vn &0.95 &0.81 \\
HD~16831 &8.96 &+0.32 &+0.17 &A5 Vp &0.16 &0.07 \\
HD~16893 &8.53 &+0.39 &+0.37 &A3~V &0.30 &0.26 \\
HD~16907 &8.39: &+0.13 &--0.07 &B9.5~V &0.16 &$\cdots$ \\
HD~17179 &7.92: &+0.26: &--0.39: &B3~Vn &0.46 &$\cdots$ \\
HD~17327 &7.49 &+0.35 &--0.03 &B8~III &0.44 &0.40 \\
HD~17327b &10.33 &+0.51 &$\cdots$ &A2~Vn &0.45 &$\cdots$ \\
HD~17443 &8.74 &+0.30 &+0.13 &B9~V &0.36 &0.31 \\
HD~17706 &8.45 &+0.38 &--0.18 &B5~IV &0.54 &0.43 \\
HD~17856 &8.71 &+0.31 &+0.18 &B9.5~Vn &0.34 &0.25 \\
HD~17857 &7.75 &+0.79 &+0.03 &B8~Ib &0.82 &$\cdots$ \\
HD~17929 &7.84 &+0.29 &--0.15 &B9~III &0.35 &0.19 \\
HD~17982 &8.07 &+0.43 &+0.39 &A1~V &0.40 &0.34 \\
HD~19065 &5.90 &--0.02 &--0.13 &B9~V &0.04 &0.05 \\
HD~19856 &8.85 &+0.21 &--0.25 &B6~III &0.35 &0.34 \\
HD~20226 &8.62 &+0.25 &--0.17 &B7~IV &0.37 &0.30 \\
HD~20336 &4.86 &--0.13 &--0.75 &B2~Vne &0.11 &0.11 \\
HD~20566 &8.08 &+0.38 &--0.24 &B3~Vne &0.58 &0.47 \\
HD~20710 &7.61 &+0.08 &--0.19 &B8~V &0.17 &0.11 \\
HD~21267 &8.00 &+0.00 &--0.29 &B7.5~V &0.11 &0.09 \\
HD~21725 &9.12 &+0.21 &+0.10 &B9.5~V &0.24 &0.17 \\
HD~21930 &8.44 &+0.19 &+0.02 &B9~VmA3 &0.25 &0.20 \\
HD~23475 &4.47 &+1.88 &+2.13 &M2~IIa &0.27 &$\cdots$ \\
BD+68$^{\circ}$193 &9.48 &+0.32 &+0.02 &B9.5~IV &0.35 &$\cdots$ \\
BD+68$^{\circ}$194 &10.06 &+0.39 &+0.22 &A1~V &0.36 &$\cdots$ \\
BD+68$^{\circ}$195 &10.20 &+0.31 &+0.01 &B9~III-IV &0.37 &0.31 \\
BD+68$^{\circ}$201 &9.68 &+0.21 &--0.06 &B9~III-IV &0.27 &0.24 \\
BD+68$^{\circ}$203 &10.22 &+0.25 &+0.13 &B9~V &0.28 &0.20 \\
TE~1 &11.03 &+1.56 &+1.37 &G5~III &0.66 &$\cdots$ \\
TE~2 &12.60 &+0.74 &+0.27 &F3~V &0.33 &0.28 \\
TE~3 &11.06 &+0.47 &+0.37 &A0~V &0.47 &0.37 \\
TE~5 &10.70 &+0.75 &+0.07 &B5~II &0.89 &0.78 \\
TE~6 &12.51 &+0.85 &+0.71 &A0~V &0.85 &0.71 \\
TE~7 &11.99 &+0.75 &+0.65 &A0~VmA5 &0.75 &0.65 \\
TE~8 &13.80 &+0.98 &+0.59 &F2~V &0.63 &0.58 \\
TE~10 &8.15 &+1.49 &+1.59 &K0~III &0.48 &$\cdots$ \\
FM~C &11.28 &+0.34 &+0.32 &A2~V &0.28 &0.23 \\
FM~G &11.29 &+0.63 &+0.36 &A8~V &0.39 &0.30 \\
FM~K &10.91 &+0.50 &+0.36 &B9.5~V &0.53 &0.43 \\
T4313-918 &10.19 &+0.42 &$\cdots$ &B6~V &0.56 &$\cdots$ \\
T4313-863 &10.53 &+0.22 &$\cdots$ &B9.5~IV &0.25 &$\cdots$ \\
\hline
\end{tabular}
\end{center}
HD~16907 = eclipsing binary TW Cas. \\
HD~17179 = V793 Cas, double-lined spectroscopic binary.
\end{table}

For the present study we have combined the original {\it UBV} measures by \citet{hv71}, \cite{ah72}, \citet{fm76}, and \citet{te84} with {\it UBV} photometry obtained by transforming the available Str\"{o}mgren photometry \citep[][and Forbes, unpublished]{fm76,sc78} using the relationships of \citet{tu90}. For a few fainter objects in the sample there are {\it BV} data from the {\it Hipparcos/Tycho} database. The data are summarized in Table~\ref{tab1} along with MK spectral types for the same stars obtained from the literature \citep[see][]{hv71}, the studies of \citet{ah72} and \citet{te85}, or from new CCD spectra obtained at dispersions of 60 \AA$\!$ mm$^{-1}$ and 120 \AA$\!$ mm$^{-1}$ in November and December 2011 using the 1.8m Plaskett telescope of the Dominion Astrophysical Observatory. Stars designated as ``FM'' are numbered by \citet{fm76}, those as ``TE'' by \citet{te84}, and, for completeness, stars designated as ``T'' are numbered from the {\it Tycho} Catalogue along with their {\it BV} data. The data for HD~23475 are from the literature \citep{ra68}, although, like many of the stars in Table~\ref{tab1}, it appears to be unrelated to SU Cas according to the present study.

\begin{figure}[!t]
\begin{center}
\includegraphics[width=6.5cm]{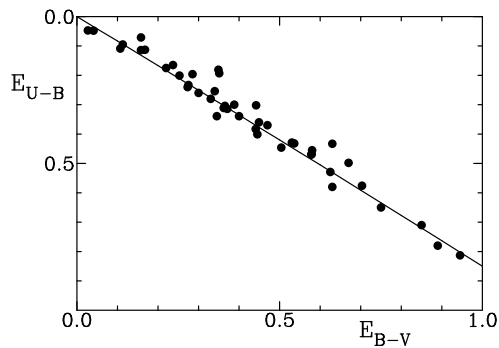}
\end{center}
\caption{\small{The reddening relation for stars near SU Cas derived from spectroscopic colour excesses. The plotted relation is described by $E_{U-B}/E_{B-V}=0.83+0.02E_{B-V}$.}}
\label{fig2}
\end{figure}

Colour excesses, $E_{B-V}$ and $E_{U-B}$, were derived for stars in Table~\ref{tab1} with reference to an unpublished set of intrinsic colours for early-type stars established by the lead author through a melding of published tables by \citet{jo66} and \citet{fi70}, subsequently confirmed through applications to stars in a variety of Galactic star fields \citep[e.g.,][]{tu89}. The resulting values are plotted in Fig.~\ref{fig2}. Intrinsic {\it (U--B)}$_0$ colours for post-main-sequence late B-type stars \citep[e.g.,][]{jo66,fi70} are particularly uncertain, which may account for some of the scatter in the diagram, notably scatter towards systematically small values of $E_{U-B}$. Excess Balmer continuum emission can also account for such effects in Be stars \citep{sr76}. As noted by \citet{tu89}, the colour excess data for stars in constrained regions of sky otherwise describe reddening lines with a typical curvature term of $+0.02$, but with a slope ranging between extremes of +0.55 and +0.85, depending upon the line of sight along which one is viewing. The data for the majority of stars near SU Cas closely fit a reddening relation described by $E_{U-B}/E_{B-V}=0.83+0.02E_{B-V}$, provided that the three most deviant objects are omitted, similar in slope to what was found for Cyg OB2 \citep{tu89}. That relation was adopted for subsequent analysis. Reddening slope and {\it R}-value are closely correlated for nearby dust clouds in the first and second Galactic quadrants \citep{tu89,tu96}, and in the present case imply a value of $R=A_V/E_{B-V}=2.8$, which was also adopted in the analysis.

The last result is important for establishing the distance to Alessi 95, so was examined carefully. At first glance it appears to conflict with the results of an earlier survey of the region by \citet{tu76b}, who found that the typical reddening law for star clusters near the Galactic longitude of SU Cas ($\ell=133\degr.4675$) had a slope $E_{U-B}/E_{B-V}$ close to 0.75--0.76, with values of $R$ averaging 3.0--3.1. In the study of \citet{te84} a value of $R=3.1$ was, in fact, adopted for the reddening corrections. Yet a re-examination of the variable-extinction data in that study (their Fig.~4) indicates that an extinction ratio of $R=2.8$ provides a much better fit to the observations than is the case for the larger value. Since the large reddening slope evident for stars near SU Cas (Fig.~\ref{fig2}) cannot be reduced to 0.75--0.76, it appears that our adoption of $R=2.8$ for those objects is a reasonable assumption. The origin of the difference relative to the \citet{tu76b} results may lie in the location of SU Cas well away from the Galactic plane ($b=+8\degr.5195$), where a localized pocket of dust has different properties from that for dust lying closer to the Galactic equator.

The reddening and distance to SU Cas and its associated dust cloud were established previously by \citet{tu84} and \citet{te85} using star counts and derived reddenings for stars near the Cepheid with {\it UBV} photometry, in conjunction with a technique developed by \citet{hs81} tied to star counts for totally opaque dust globules. Given the new information that the Cepheid lies in the core of a previously-unnoticed open cluster, the use of star counts may no longer be appropriate for the study of extinction and distance, leaving the question of the reddening for the Cepheid and cluster open to further examination.

\begin{figure}[!t]
\begin{center}
\includegraphics[width=6.5cm]{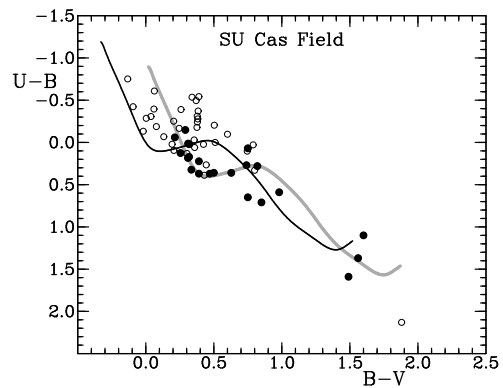}
\end{center}
\caption{\small{Colour-colour diagram for stars in the SU Cas field, with stars lying within $\sim 1^{\circ}$ of the Cepheid plotted with filled symbols. The black curve is the intrinsic relation for dwarfs, while the thick gray curve represents the intrinsic relation reddened by $E_{B-V}=0.35 \pm0.02$.}}
\label{fig3}
\end{figure}

\begin{table*} \footnotesize
\caption[]{Abbey Ridge Observatory {\it BV} observations for stars near SU Cas.}
\label{tab2}
\begin{center}
\begin{tabular}{@{\extracolsep{+0.5mm}}ccccccccccc}
\hline
Star &RA(2000) &DEC(2000) &{\it V} &{\it B--V} & &Star &RA(2000) &DEC(2000) &{\it V} &{\it B--V} \\
\hline
1 &42.6032 &68.9848 &11.29 &0.59 & &33 &42.6618 &68.8036 &15.32 &2.17 \\
2 &43.5207 &68.8292 &11.94 &0.58 & &34 &42.5644 &68.7939 &15.41 &1.87 \\
3 &42.5635 &68.8747 &12.28 &0.62 & &35 &42.7140 &68.8649 &15.53 &2.20 \\
4 &42.5214 &68.9379 &13.15 &0.74 & &36 &42.6204 &68.8376 &15.57 &1.43 \\
5 &43.4562 &68.8101 &13.58 &0.86 & &37 &42.8804 &68.8988 &15.62 &1.39 \\
6 &42.7422 &68.9688 &13.64 &0.96 & &38 &42.9933 &68.9558 &15.67 &1.69 \\
7 &42.9819 &68.9792 &13.67 &0.92 & &39 &43.3421 &68.7724 &15.71 &2.68 \\
8 &42.7272 &68.8951 &13.75 &1.23 & &40 &42.8359 &68.8794 &15.72 &1.72 \\
9 &42.6903 &68.7487 &13.83 &0.80 & &41 &42.6393 &68.8148 &15.80 &2.02 \\
10 &43.4705 &68.8416 &13.91 &1.12 & &42 &42.5626 &68.7751 &15.82 &1.40 \\
11 &43.1982 &68.7507 &14.30 &2.18 & &43 &43.2426 &68.7531 &15.83 &1.46 \\
12 &43.4667 &68.9085 &14.43 &0.99 & &44 &42.6391 &68.9859 &15.89 &1.76 \\
13 &42.5669 &68.8139 &14.49 &0.91 & &45 &43.4950 &68.9855 &15.94 &1.39 \\
14 &43.1307 &68.8611 &14.73 &1.28 & &46 &43.0456 &68.9040 &15.99 &1.78 \\
15 &42.7000 &68.9516 &14.75 &1.07 & &47 &42.7820 &68.8385 &16.08 &1.51 \\
16 &43.4772 &68.7904 &14.79 &0.93 & &48 &43.3109 &68.8668 &16.09 &1.53 \\
17 &42.6767 &68.8458 &14.81 &1.02 & &49 &42.8790 &68.7889 &16.15 &2.01 \\
18 &43.5128 &68.9190 &14.90 &2.00 & &50 &43.2876 &68.8010 &16.20 &1.69 \\
19 &43.1200 &68.9134 &14.91 &1.33 & &51 &42.9461 &68.9348 &16.22 &2.06 \\
20 &43.0688 &68.8097 &14.91 &1.14 & &52 &42.5290 &68.8124 &16.29 &1.79 \\
21 &42.5517 &68.8297 &15.00 &0.88 & &53 &43.1040 &68.9229 &16.32 &2.05 \\
22 &43.2084 &68.9804 &15.04 &1.47 & &54 &43.3733 &68.8587 &16.42 &1.49 \\
23 &43.2299 &68.8139 &15.04 &2.99 & &55 &42.6019 &68.9187 &16.43 &1.80 \\
24 &43.5362 &68.8662 &15.07 &2.29 & &56 &43.5272 &68.8812 &16.44 &1.66 \\
25 &43.4419 &68.7818 &15.08 &1.19 & &57 &42.6220 &68.9404 &16.49 &1.71 \\
26 &42.5198 &68.7896 &15.12 &1.02 & &58 &42.8224 &68.8314 &16.49 &1.80 \\
27 &42.7201 &68.7484 &15.13 &2.29 & &59 &42.6265 &68.9563 &16.51 &1.55 \\
28 &43.3182 &68.7498 &15.13 &1.08 & &60 &42.7998 &68.9394 &16.60 &1.67 \\
29 &42.7834 &68.9211 &15.18 &1.38 & &61 &43.0472 &68.7603 &16.77 &1.98 \\
30 &43.5239 &68.9916 &15.21 &0.99 & &62 &42.6304 &68.7652 &16.84 &2.53 \\
31 &43.0970 &68.7965 &15.23 &1.33 & &63 &42.6872 &68.7974 &16.94 &1.71 \\
32 &42.8776 &68.8835 &15.24 &1.05 & &64 &42.9295 &68.8324 &17.01 &1.59 \\
\hline
\end{tabular}
\end{center}
ARO~3 = TYC~4313-355, Sp.T. = F1~V.$\;\;$ ARO~4, Sp.T. = A3~V.
\end{table*}

The colours of stars in the SU Cas field are plotted in Fig.~\ref{fig3}, with filled circles used to denote stars lying in relatively close proximity to SU Cas. Reddenings have also been derived for four additional stars lying near the Cepheid from {\it BV} data and spectral types by \citet{ah72} along with two stars from Table~\ref{tab2} below, four of the six appearing to lie at similar distances to SU Cas. The derived mean reddening for the collection of 13 stars lying in close proximity to the Cepheid, and not projected against an obvious dust cloud, is $E_{B-V}({\rm B0})=0.35\pm0.02$ s.e. ($\pm0.05$ s.d.), which also appears to apply to a few stars in the Table~\ref{tab1} collection (see mean reddening adopted in Fig.~\ref{fig3}). There is noticeable differential reddening in the field according to the observations, that near SU Cas being associated with the visible dust clouds around the Cepheid. The mean reddening of the central regions of Alessi 95 can be solidly established as $E_{B-V}({\rm B0})=0.35\pm0.02$ s.e., however, with $E_{B-V}=0.33\pm0.02$ s.e. inferred for the space reddening of a star with the observed colours of SU Cas \citep[see][]{fe63}.

The inferred reddening for the Cepheid is a close match to an estimate of $E_{B-V}=0.32$: established by \citet*{te87} from published spectrophotometric {\it KHG} photometry for SU Cas in conjunction with intrinsic values established from Cepheids of well-established space reddening. The reddening also agrees reasonably well with an estimate of $E_{B-V} =0.296\pm0.026$ obtained by \citet{ko08} from stellar atmosphere model fitting, in this case linked to derived intrinsic colours and effective temperatures for bright stars of little to no reddening. A reddening of $E_{B-V}=0.28$ was derived by \citet{lc07} from {\it BVI}$_c$ photometry for SU Cas linked to a calibration based on space reddenings for Cepheids, but that included the earlier estimate of $E_{B-V}=0.27\pm0.03$ s.e. by \citet{tu84}. Given that SU Cas is the shortest period Cepheid in the sample of pulsators with established space reddenings, the small offset of the \citet{lc07} reddening with the present result is presumably linked to the original underestimate of reddening for SU Cas obtained by \citet{tu84}.

\begin{figure}[!t]
\begin{center}
\includegraphics[width=6.5cm]{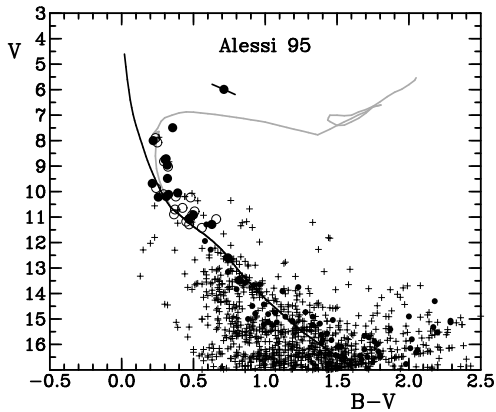}
\end{center}
\caption{\small{Uncorrected colour-magnitude diagram for Alessi 95, with photoelectric data identified by large filled circles, stars from the {\it Hipparcos/Tycho} database by open circles, stars from the ARO survey by small filled circles, stars from {\it NOMAD} lying in the inner $30^{\prime}$ of the cluster by small plus signs, and SU Cas by a filled circle with bars to indicate its range of variability. The black relation is the ZAMS for $V-M_V=9.11$, and a gray curve is a model isochrone for $\log t = 8.2$.}}
\label{fig4}
\end{figure}

Potential members of Alessi 95 were assembled from stars lying within the cluster boundaries: photoelectrically-observed stars (Table~\ref{tab1}), stars in the {\it Hipparcos/Tycho} database \citep{es97,vl07}, stars brighter than $V=17$ in the {\it NOMAD} database \citep{za05}, recalibrated to the Johnson {\it BV} system using faint stars from \citet{tu84} as reference standards, and stars near the Cepheid with new CCD {\it BV} observations (see below). Obvious ``ringers'' among the faint stars in {\it NOMAD} were omitted from the analysis, and the resulting data are plotted in Fig.~\ref{fig4}, which represents the colour-magnitude diagram for Alessi 95 uncorrected for differential reddening or detailed membership selection. Included are a best-fitting zero-age main sequence (ZAMS, see below) and a model isochrone from \citet*{me93} for $\log t=8.2$, which appears to fit the data for cluster stars and the Cepheid SU Cas reasonably well. Note that late-type dwarfs are only encountered in this direction at the distance of Alessi 95 and beyond, so the cluster cannot be less distant than implied from the ZAMS fit, for example at the distance of $258\pm3$ pc derived for the foreground dust complex \citep{te84}.

CCD observations of the central $15^{\prime} \times 22^{\prime}$ region surrounding SU Cas were made through Johnson system {\it BV} filters in September 2011 with the SBIG ST8XME camera on the Celestron 35-cm telescope of the robotic Abbey Ridge Observatory \citep[see][]{la08}. The observations were calibrated using previously published photoelectric {\it UBV} photometry for stars in the field \citep[see][and this paper]{tu84,te84}, and are included in Fig.~\ref{fig4}.

The reddening for individual stars was established using the reddening relation found from the spectroscopic observations (Fig.~\ref{fig2}) through standard dereddening techniques \citep[see][]{tu76a,tu76b}. For stars with {\it BV} data only, colour excesses were inferred from the spatial trend of reddening across the field, except for those objects closely associated with the opaque dust clouds near the Cepheid. In all cases the B0-star reddenings averaged for spatially-adjacent stars were adopted for individually dereddened stars, then adjusted for the colour dependence of reddening to that appropriate for the inferred intrinsic colour of each object \citep[see][]{fe63}. In a few cases it was possible to infer an independent reddening from the spectral classification for the star. Each star was corrected for extinction using its inferred B0-star reddening in conjunction with the adopted value of $R=2.8$.

\begin{figure}[!t]
\begin{center}
\includegraphics[width=6.5cm]{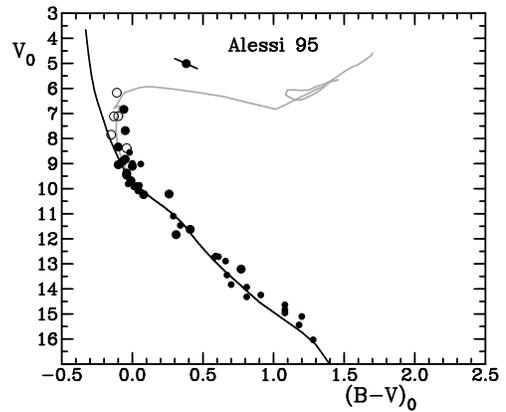}
\end{center}
\caption{\small{Reddening and extinction-free oolour-magnitude diagram for Alessi 95, with photoelectric data identified by large filled circles, stars from the ARO survey and {\it Hipparcos/Tycho} database by small filled circles, and SU Cas as in Fig.~\ref{fig4}. Open circles denote the 6 stars from Table~\ref{tab1} that may be outlying cluster members. The black relation is the ZAMS for $V_0-M_V=8.16$, and the gray curve is a model isochrone for $\log t = 8.2$.}}
\label{fig5}
\end{figure}

The resulting data are plotted in Fig.~\ref{fig5} along with similarly-derived data for six stars in Table~\ref{tab1} that appear to share comparable space motions and parallaxes with SU Cas. Observed radial velocities \citep[see][]{te84,te85}, proper motions \citep{vl07}, and parallaxes \citep{vl07} were used to identify only {\it bona fide} potential outlying cluster members on the basis of similarity of the values to those for SU Cas.

\begin{deluxetable}{cccc}
\tablewidth{0pt}
\tabletypesize{\small}
\tablecaption{\small{Hipparcos parallax data for members and potential outlying members of Alessi 95}}
\tablehead{\colhead{Hipparcos} &\colhead{Star} &\colhead{$\pi_{\rm abs}$} &\colhead{$\pm \sigma_{\pi}$} \\
&\colhead{} & \colhead{(mas)} &\colhead{(mas)}}
\startdata
11633 &BD+68$^{\circ}$~170 &2.31 &1.46 \\
12434 &HD~16228 &2.80 &1.23 \\
12567 &HD~16393 &3.03 &0.52 \\
12924 &HD~16841 &3.08 &1.01 \\
13138 &BD+70$^{\circ}$~205 &1.04 &1.02 \\
13208 &HD~17327 &0.98 &0.77 \\
13219 &BD+69$^{\circ}$~181 &1.82 &1.20 \\
13367 &SU~Cas &2.53 &0.32 \\
13465 &BD+68$^{\circ}$~201 &0.89 &1.34 \\
13595 &BD+69$^{\circ}$~186 &0.82 &1.47 \\
13728 &HD~17929 &2.84 &0.64 \\
13956 &BD+69$^{\circ}$~189 &2.36 &1.40 \\
14442 &BD+70$^{\circ}$~224 &0.60 &1.10 \\
14483 &BD+67$^{\circ}$~243 &--0.97 &1.99 \\
14714 &HD~19287 &1.02 &1.32 \\
15138 &HD~19856 &1.13 &1.09 \\
15959 &HD~20710 &4.19 &0.61 \\
16633 &HD~21725 &0.86 &1.25 \\

\enddata
\tablenotetext{1}{{\scriptsize BD+68$^{\circ}$~201 = FM-A.}}
\label{tab3}
\end{deluxetable}
The reddening for individual members of Alessi 95 is well enough established for 26 likely ZAMS members of the cluster to derive a mean intrinsic distance modulus of $V_0-M_V=8.16\pm0.04$ s.e. ($\pm0.21$ s.d.). The zero-age main sequence (ZAMS) adopted here is that of \citet{tu76a,tu79}, and the fit corresponds to a distance of $d=429\pm8$ pc, with the cited uncertainty representing the standard error of the mean. A model isochrone for $\log t = 8.2$ from \citet{me93} fits the data reasonably well, with a likely uncertainty in $\log t$ no larger than $\pm 0.1$. It is possible to use alternate evolutionary isochrones from the literature, but they do not match the adopted ZAMS nearly as well. The isochrone fit for Alessi 95 is not ideal for SU Cas, but that problem could be resolved by accounting for the opacity effects of CNO mixing in the envelopes of post-supergiant stars \citep[see][]{xl04}. A more typical solution to the problem of a compressed blue loop for core helium-burning stars is to adopt a metallicity for the isochrone that is smaller than the solar value, but that does not appear to be justified in the case of Alessi 95, given that the derived metallicity of SU Cas (and two associated stars) from stellar atmosphere models is close to solar ($\langle$[Fe/H]$\rangle$ = --0.12, +0.02, --0.01, and +0.06 according to \citet{ko96,us01,an02,lu08}, respectively).

Three of the bright members of Alessi 95, as well as 15 other stars lying within a few degrees of SU Cas sharing similar space motions (proper motions and in a few cases radial velocities) with the Cepheid and with colours and magnitudes consistent with the $\log t=8.2$ isochrone in Fig.~\ref{fig4}, are catalogued in the {\it Hipparcos} catalogue \citep{vl07}. The stars are collected in Table~\ref{tab3}, along with their cited parallaxes and uncertainties. The weighted mean parallax for the group is $\pi_{\rm abs}=2.38 \pm0.19$ mas, corresponding to a distance of $420 \pm33$ pc. An attempt was also made to include the less precise parallaxes from the {\it Tycho} catalogue \citep{es97} in the result. A further fifteen stars were considered in such fashion, but the resulting mean parallax and distance remained unaffected, since the extra stars add negligible weight to the overall solution because of their large parallax uncertainties. In both cases the parallax solution is in excellent agreement with the cluster distance of $429 \pm8$ pc derived from ZAMS fitting.

\begin{figure}[!t]
\begin{center}
\includegraphics[width=6cm]{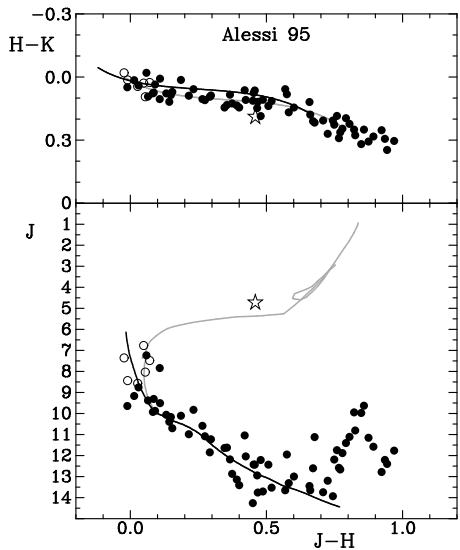}
\end{center}
\caption{\small{{\it JHK}$_s$ colour-colour diagram (upper) and colour-magnitude diagram (lower) for stars in Alessi 95 (filled circles) and outlying stars (open circles), indicated to be potential cluster members from proper motion data. SU Cas is denoted by a star symbol. The black relation in the upper diagram is the intrinsic relation for dwarfs, while the gray relation corresponds to a reddening of $E_{B-V}=0.35$. The black relation in the lower diagram is the ZAMS for $V_0-M_V=8.16$, and the gray curve is a model isochrone for $\log t = 8.2$.}}
\label{fig6}
\end{figure}

A further consistency check on the results was made using {\it JHK}$_s$ photometry \citep{cu03} from the Two Micron All Sky Survey \citep[2MASS,][]{sk06} for Galactic star fields. Stars lying within $50^{\prime}$ of the centre of Alessi 95 with proper motions similar to that of SU Cas are plotted in colour-colour and colour-magnitude diagrams in Fig.~\ref{fig6} in the manner adopted by \citet{tu11}. Also included are similar data for the stars from Table~\ref{tab1} and Fig.~\ref{fig5} selected on the basis of comparable space motion with SU Cas. Most of the scatter in the observations can be attributed to photometric uncertainties, yet the data are a reasonably close match to the results inferred from the {\it UBV} analysis, namely the implied reddening and distance modulus.

A program was also completed in January 2012 to obtain {\it JHK}$_s$ photometry of greater precision for the field of Alessi 95 using queue observing with the near-infrared imager (CPAPIR) of l'Observatoire du Mont-M\'{e}gantic \citep[OMM][]{ar10}. The details of that study, to be presented elsewhere, provide much stronger confirmation of the optical results than is the case for the less precise 2MASS data, and also provide information on the extreme lower end of the cluster main sequence. 

The nuclear and coronal radii for Alessi 95 implied by its distance derived from ZAMS fitting are 3 pc and $6\frac{1}{4}$ pc, respectively. Both values are reasonable, although a larger tidal radius is expected, which would explain the presence of outlying cluster members a few degrees from SU Cas.

\section{{\rm \footnotesize SU CAS AS A MEMBER OF ALESSI 95}}

In the studies by \citet{sc78} and \citet{te84} of stars associated with the dust complexes near SU Cas it was noted that the dust clouds lie at different distances. The two main clouds at the declination of SU Cas and south of it are at distances of $258\pm3$ pc and $401\pm38$ pc according to the \citet{te84} study. SU Cas was assumed to lie at the distance of the former, according to a perceived connection on Palomar Observatory Sky Survey (POSS) plates of the SU Cas dust complex with that associated with HD~16893 and HD~17443, which were calculated to be 258 pc distant.

The deeper POSS images used for the production of Fig.~\ref{fig1} indicate that such an assumption cannot be correct. The reflection nebulae illuminated by SU Cas that lie south of the Cepheid display no association with the opaque dust cloud east of it that continues further to the northeast. Strands of the same opaque cloud are seen to the west and south of SU Cas, connecting with the clouds illuminated by HD~16893 and HD~17443, but the main cloud east of SU Cas displays no evidence for associated reflection nebulosity, so must lie foreground to the Cepheid. The true geometry of the stars and dust is revealed by those members of Alessi 95 that, from their large reddenings, must be viewed {\it through} the dust extinction of the main cloud. SU Cas and Alessi 95 therefore lie beyond the main dust complex located at 258 pc. Indeed, the derived distance of $429\pm8$ pc to Alessi 95 agrees closely with the previous estimate \citep{te84} for the distance to the further dust complex in the field.

The derived luminosity for SU Cas as a member of Alessi 95, including the uncertainty in its field reddening, is $\langle M_V \rangle = -3.15 \pm0.07$ s.e., where the Cepheid's magnitude \citep{be07} has been adjusted for contamination by an unseen B-type companion 4.2 magnitudes fainter \citep{ef87}. The result is more than a magnitude more luminous than expected for a classical Cepheid with a mean period of $P=1.949322$ days, as used in studies of the star's period changes \citep*{be97,be03,te06}. The Cepheid must therefore be an overtone pulsator, as argued previously \citep[e.g.,][]{gi76,gi82}. According to the empirical relationship between fundamental mode and first overtone pulsation periods established for double-mode Cepheids by \citet{sz88}, the undetected mean period for fundamental mode pulsation in SU Cas must be $2.754776$ days. 

An independent distance estimate for SU Cas is possible from its radius and inferred mean effective temperature, using the infrared surface brightness variant of the Baade-Wesselink method, as noted by \citet{st11}. The estimated distance to SU Cas in that study is $418 \pm12$ pc for a pulsation period of $1^{\rm d}.95$. With the revised space reddening found here for SU Cas and a local ratio of total-to-selective extinction of $R=2.8$, the derived distance becomes $414 \pm12$ pc for fundamental mode pulsation ($P=2^{\rm d}.75$). The distance estimates are not completely independent, but the methodologies are. The close agreement in the pulsation parallax, trigonometric parallax, and cluster parallax estimates for the distance to Alessi 95 and SU Cas --- $414\pm12$ pc, $420\pm33$ pc, and $429\pm8$ pc, respectively, all of which agree to within their derived uncertainties --- provides strong confirmation of their validity.

The implied mean radius of SU Cas according to the Cepheid period-radius relation of \citet{te10}, which is tied to the almost identical slopes for the Cepheid period-radius relation in studies by \citet*{gi89}, \citet{ls95}, \citet*{gi98}, and \citet{tb02}, is 25.0 $R_{\odot}$. By comparison, the surface brightness technique yields a mean radius of $28.0 \pm0.8$ $R_{\odot}$, while an independent Baade-Wesselink analysis using infrared colours by \cite{mi99} produced an estimate of $33.0\pm1.1$ $R_{\odot}$. The last study also summarizes previous estimates for the mean radius of SU Cas derived from variants of the Baade-Wesselink method, most of which lie in the range 30--45 $R_{\odot}$. Systematic effects cannot be discounted, given the small amplitude of the light variations in SU Cas and the fact that \citet{mi99} adopted a projection factor of $p=1.39$ in their study. \citet{lj09} find a value of $p=1.277$ to be more suitable for SU Cas, which would reduce the \citet{mi99} value to $30.3\pm1.0$ $R_{\odot}$. The {\it VJK} photometry used by \citet{lj09} leads to a mean radius of $28.5$ $R_{\odot}$ for SU Cas, while \citet{tb02} obtained a radius of $19.0\pm0.7$ $R_{\odot}$ using a Baade-Wesselink analysis tied to {\it KHG} photometry. The last three estimates are closer to the value predicted by the Cepheid period-radius relation, and it may be that the star's small pulsation amplitude and contamination by its B-type companion limit further improvement. 

With the period of fundamental mode pulsation in SU Cas indicated by its membership in Alessi 95, one can estimate independently the age of the Cepheid from existing period-age relations. A model-based relationship derived by \citet{bo05} yields an age of $\log t = 8.0$ for a solar metallicity Cepheid with $P=2^{\rm d}.75$, while a relationship by \citet{ee98} calibrated by isochrone fits to open clusters yields a similar age of $\log t = 8.3$. Both cited relationships display a dispersion in $\log t$ of $\pm0.1$, so are consistent with the cluster age inferred from the model isochrones of \citet{me93}. Additionally, the implied age of $\log t = 8.2$ for SU Cas from its membership in Alessi 95 provides a key point at the short-period end of a semi-empirical period-age relationship for cluster Cepheids developed by \citet{tu12}, where again the dispersion is no larger than $\pm 0.1$ in $\log t$ and the relationship is generated by isochrone fits to clusters containing Galactic Cepheids.  

The luminosity of SU Cas inferred from cluster membership, namely $\langle M_V \rangle = -3.15 \pm0.07$, can be compared with a value of $\langle M_V \rangle = -2.77$ predicted from its inferred radius and effective temperature via the methodology described by \citet{tb02} and \citet{te10}. The $0^{\rm m}.38$ offset from the empirical estimate is comparable to the offsets observed for other cluster Cepheids \citep{tu10}, and can be partially explained by the nature of SU Cas: a small-amplitude Cepheid lying near the high temperature side of the instability strip, as also argued by its overtone pulsation. SU Cas has an unseen B9.5~V companion detected by IUE \citep{ef87}, and the luminosity inferred for the Cepheid according to the implied magnitude difference between SU Cas and the B-star in ultraviolet spectra is $M_V=-3.2 \pm0.1$ \citep{ef87}, closely coincident with the result from cluster membership. The companion, which may also be an Ap star \citep{tu03}, has otherwise only a small effect on the overall visual brightness and colours of SU Cas.

Confirmation of the location of SU Cas towards the blue edge of the Cepheid instability strip is provided by its observed rate of period change of +0.024 s yr$^{-1}$ \citep{be97,be03,te06}. That implies a rate of evolution for the Cepheid that is more than twice as rapid as what is typical of fundamental mode pulsators with periods of $\sim 2^{\rm d}.75$ near the centre of the instability strip \citep{te06}. The consistency in the implied parameters for SU Cas found from such diverse observational material provides further validation of the present results.

\section{{\rm \footnotesize DISCUSSION}}

This first detailed photometric and spectroscopic study of Alessi 95, the sparse open cluster surrounding the Cepheid SU Cas, produces empirical estimates for the reddening and luminosity of a {\it bona fide} Cepheid calibrator. Estimates for the distance to SU Cas and Alessi 95 from ZAMS fitting, {\it Hipparcos} parallaxes \citep{vl07}, and the Cepheid's pulsation parallax \citep[][adjusted here]{st11} agree closely to within a few percent, to our knowledge the first such instance of a tight consensus in distance estimates by these diverse methods. Future improvement to the cluster ZAMS fit might be possible, for example through a more detailed analysis that accounts implicitly for the metallicity of SU Cas and Alessi 95 members. That affects ZAMS fitting through small offsets in the zero-point, which is currently calibrated for solar metallicity stars. Given present knowledge of the near-solar chemical composition for Alessi 95 stars \citep{us01}, however, only a very minor improvement would be expected.

The inferred field reddening of $E_{B-V}=0.33\pm0.02$ found here for SU Cas agrees reasonably well with other independent estimates \citep{te87,lc07,ko08}, and the effective temperature of 6620K for SU Cas inferred from its colour via the semi-empirical technique described by \citet{tb02} and \citet{te10} is only slightly greater than the values obtained by \citet{us01} from stellar atmosphere models. Any discrepancies between the present results and those of previous studies appear to be minor, except for the inferred luminosity of $\langle M_V \rangle = -3.15 \pm0.07$ s.e., which differs from that found in the earlier study by \citet{te84}, primarily because of an erroneous selection of reflection nebulosity stars associated with the Cepheid in that paper. Future work may improve the situation, since {\it Hipparcos} parallaxes, pulsation parallaxes, and cluster parallaxes are all susceptible to systematic effects that are the subject of ongoing study. It should be evident, however, that there is no longer a need to rely on the Cepheid's membership in Cas R2 to derive its intrinsic parameters \citep[e.g.,][]{te84,us01,tu10}.

\subsection*{{\rm \scriptsize ACKNOWLEDGEMENTS}}
\scriptsize{This publication makes use of data products from the Two Micron All Sky Survey, which is a joint project of the University of Massachusetts and the Infrared Processing and Analysis Center/California Institute of Technology, funded by the National Aeronautics and Space Administration and the National Science Foundation. The authors gratefully acknowledge the use of data products from the SAO/NASA Astrophysics Data System (ADS) in this study, and are indebted to the Dominion Astrophysical Observatory for the generous allotment of observing time for the new spectroscopic results presented here and to Noel Carboni for the creation of Fig.~\ref{fig1}. We also thank the referee, Clifton Laney, for helpful suggestions on the original manuscript. WPG gratefully acknowledges support from the Chilean Center for Astrophysics FONDAP 15010003 and the BASAL Centro de Astrofisica y Tecnologias Afines (CATA) PFB-06/2007.}

\end{document}